\documentclass[aps,twocolumn,showpacs,amsmath,amssymb]{revtex4}

\usepackage{graphicx}

\newcommand{\be}{\begin{eqnarray}}
\newcommand{\ee}{\end{eqnarray}}

\newcommand{\nn}{\nonumber}

\begin{document} 
\title{A geometric phase gate without dynamical phases} 
\author{R. G. Unanyan and M. Fleischhauer}
\affiliation{Fachbereich Physik der Universit\"{a}t Kaiserslautern, 67653\\
Kaiserslautern, Germany}
\date{\today} 

%%%%%%%%%%%%%%%%%%%%%%%%%%%%%%%%%%%%%%%%%%%%%%%%%%%%%%%%%%%%%%%%%%%%%%%%%%
%%%%%%%%%%%%%%%%%%%%%%%%%%%%%%%%%%%%%%%%%%%%%%%%%%%%%%%%%%%%%%%%%%%%%%%%%%
 
\begin{abstract} 
A general scheme for an adiabatic geometric phase 
gate is proposed which is maximally robust
against parameter fluctuations. While in systems with SU(2) symmetry
geometric phases are usually accompanied by dynamical phases and are thus not 
robust, we show that in the more general case of a SU(2)$\otimes$ SU(2)
symmetry it is possible to obtain a non-vanishing geometric phase without
dynamical contributions. 
The scheme is illustrated for a phase gate using two systems 
with dipole-dipole interactions in external laser fields
which form an effective four-level system.
\end{abstract} 
 

%%%%%%%%%%%%%%%%%%%%%%%%%%%%%%%%%%%%%%%%%%%%%%%%%%%%%%%%%%%%%%%%%%%%%%%%%%
%%%%%%%%%%%%%%%%%%%%%%%%%%%%%%%%%%%%%%%%%%%%%%%%%%%%%%%%%%%%%%%%%%%%%%%%%%
 
\maketitle

%%%%%%%%%%%%%%%%%%%%%%%%%%%%%%%%%%%%%%%%%%%%%%%%%%%%%%%%%%%%%%%%%%%%%%%%%%
%%%%%%%%%%%%%%%%%%%%%%%%%%%%%%%%%%%%%%%%%%%%%%%%%%%%%%%%%%%%%%%%%%%%%%%%%%

%\section{Introduction}

%%%%%%%%%%%%%%%%%%%%%%%%%%%%%%%%%%%%%%%%%%%%%%%%%%%%%%%%%%%%%%%%%%%%%%%%%%%
%%%%%%%%%%%%%%%%%%%%%%%%%%%%%%%%%%%%%%%%%%%%%%%%%%%%%%%%%%%%%%%%%%%%%%%%%%

A major challenge for the practical implementation of fault-tolerant
scalable quantum computing 
\cite{Ekert}
is the requirement of single-bit and quantum gate 
operations approaching unity fidelity up to one part in $10^4$ 
\cite{error-correction}.
As any unitary operation on physical qubits involves 
the interaction with external systems, this implies either a very 
precise control or a clever design  of these systems to make them
insensitive to small  variations of 
external parameters. Since a universal set of qubit
operations includes arbitrary single-bit rotations, the qubit system
must depend sensitively on at least one continuous external parameter, e.g. 
some laser phase. It is therefore clear that not 
all elements of quantum computation can be made robust with respect to
external parameters. However 
apart from single-bit phase rotations, this should be possible.
This has lead to several proposals for 
adiabatic quantum computation \cite{Vedral-tutorial} 
using geometric or Berry phases \cite{Berry}. 
The choice of {\it adiabatic} evolutions removes the sensitivity 
to many external parameters at the expense of slower
operations and {\it geometric phases} depend only on the 
geometry of the path followed in parameter space 
and are thus also tolerant toward parameter fluctuations. 
However non-vanishing geometric phases are in general associated
with non-vanishing dynamical phases that need to be
compensated in some way, which destroys the robustness.
For example in the proposal of Jones et al. 
for geometric quantum computation in nuclear magnetic resonance, 
dynamical phases are compensated by spin echo techniques 
\cite{Jones-Nature-2000}. In the ion-trap system of Duan et al. 
\cite{Duan-Science-2001} dynamical phase shifts arise due to
unavoidable ac-Stark shifts which have the same strength as the
couplings used for the implementation of the phase gate.
Also in the recent proposal of Garcia-Ripoll and Cirac 
\cite{Garcia-Ripoll-PRL-2003} to perform single and two-bit operations
with unknown interaction parameters, dynamical phases arise in
the individual steps of the operation and the interaction
parameters must be controlled over the whole cycle in order for
these phase to add up to zero.

We here show that it is possible to
obtain a nonzero geometric phase under conditions of exactly vanishing
dynamical contributions at all times when going from a SU(2) system to
a case with SU(2) $\otimes$ SU(2) symmetry.
The simplest nontrivial representation of the 
SU(2) $\otimes$ SU(2) symmetry is a four-level scheme
with a tripod coherent coupling 
\cite{Unanyan-OptComm-1998,Unanyan-PRA-1999,Duan-Science-2001}.
An appropriate adiabatic rotation in parameter space creates
phase shifts of $\pi$ or $\pi/2$. This approach is then
applied to two systems with dipole-dipole interaction
to construct a phase gate.

%%%%%%%%%%%%%%%%%%%%%%%%%%%%%%%%%%%%%%%%%%%%%%%%%%%%%%%%%%%%%%%%%%%%%%%%%%
%%%%%%%%%%%%%%%%%%%%%%%%%%%%%%%%%%%%%%%%%%%%%%%%%%%%%%%%%%%%%%%%%%%%%%%%%%

%\section{geometrical and dynamical phases}

%%%%%%%%%%%%%%%%%%%%%%%%%%%%%%%%%%%%%%%%%%%%%%%%%%%%%%%%%%%%%%%%%%%%%%%%%%
%%%%%%%%%%%%%%%%%%%%%%%%%%%%%%%%%%%%%%%%%%%%%%%%%%%%%%%%%%%%%%%%%%%%%%%%%%

\

When a system undergoes a {\it cyclic evolution} the wave function of the
system remembers its motion in the form of a phase factor. As
first noted by Berry \cite{Berry}, one can distinguish two contributions to
the phase acquired: a dynamical part and a
geometrical part. Berry showed that when the Hamiltonian of the system
depends on a set of parameters which evolve {\it adiabatically} along a
closed curve in the parameter space, then the state vector corresponding to
a simple non-degenerate eigenvalue develops a phase which depends only on this
curve. 
Unlike its dynamical counterpart, the geometrical or Berry phase 
does not depend on the
duration of the interaction. The notion of the Abelian 
Berry phase was generalized to
the case of degenerate levels by Wilczek and Zee \cite{Wilczek} 
which involve non-Abelian operations
and to general
cyclic evolutions by Aharonov and Anandan \cite{Aharonov}.

\begin{figure}[ht] 
  \begin{center} 
    \includegraphics[width=3.0cm]{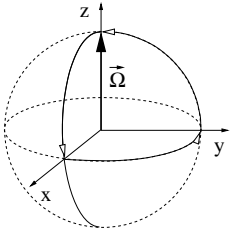} 
    \caption{Cyclic evolution of the magnetic field ${\bf\Omega}\sim
{\bf B}$ from $z$ to $x$ to $y$ and back to the $z$ direction to generate
geometric phase of $\pi/2$.
}
    \label{Fig1} 
  \end{center} 
\end{figure}

Due to their intrinsically robust nature, geometric phases are attractive
for quantum computation. Their application faces however a problem. 
As has been demonstrated by Robbins and Berry \cite{Robbins} 
for a spin-$J$ particle in a slowly and cyclically changing
magnetic field ${\bf B}$
\be
H=\hbar{\bf\Omega}\cdot \hat{\bf J},\qquad {\bf \Omega}=\mu_B {\bf B}/
\hbar\label{H1}
\ee
 the 
angular momentum state $|m\rangle$ (in the direction of ${\bf B}$)  attains
not only a geometric phase but also a dynamical one. To see this, let us
assume that ${\bf\Omega}$ makes a cyclic evolution in 
parameter space as shown in Fig.\ref{Fig1}.

%%%%%%%%%%%%%%%%%%%%%%%%%%%%%%%%%%%%%%%%%%%%%%%%%%%%%%%%%%%%%%%%%%%%%%%%%%

%%%%%%%%%%%%%%%%%%%%%%%%%%%%%%%%%%%%%%%%%%%%%%%%%%%%%%%%%%%%%%%%%%%%%%%%%%

Starting with a magnetic-field orientation 
in the $z$-direction, ${\bf\Omega}$ is successively rotated around 
the $y$, $z$, and $x$ axis by $\pm \pi/2$. The total solid angle in 
parameter space is
then $\pi/2$. In principle any other angle can be obtained, but this 
particular path can be implemented in a very robust way.
The initial state vector 
is assumed to be an eigenstate of (\ref{H1}) with eigenvalue 
$E_m=m\hbar\Omega$, i.e. 
$|\psi_0\rangle=|J,m_z=m\rangle$.
An adiabatic cyclic evolution then leads to a state 
\be
|\psi\rangle = {\rm e}^{-i(\gamma_{\rm g}+\gamma_{\rm d})} \, |\psi_0\rangle,
\ee
where $\gamma_{\rm d}$ is the dynamical phase 
\be
\gamma_{\rm d}=\int\!\! {\rm d}t\, \frac{E_m(t)}{\hbar},
\ee
and the geometrical phase can be obtained from parallel transport
$U_{\rm g}|\psi_0\rangle$
of the adiabatic eigenstate following the rotation of the magnetic field
\be
U_{\rm g} ={\rm e}^{-i\frac{\pi}{2} \hat J_x}
{\rm e}^{-i\frac{\pi}{2} \hat J_z}{\rm e}^{-i\frac{\pi}{2} 
\hat J_y}|\psi_0\rangle = {\rm e}^{-i\frac{\pi}{2} 
\hat J_z}|\psi_0\rangle,
\ee
i.e.
\be
\gamma_{\rm g}=\frac{\pi}{2}m.
\ee
One recognizes that an elimination
of the dynamical phase implies in general a vanishing geometric phase, since 
a zero-eigenstate of $H$ at $t=0$
is also an eigenstate of $U_{\rm g}$ with eigenvalue 1.

In order to have a non-vanishing geometric phase and at the same time
a vanishing dynamical one, the state $|\psi_0\rangle$ must be an eigenstate
of the initial Hamiltonian with eigenvalue zero but should not be an
eigenstate of $U_{\rm g}$ with eigenvalue unity. This can be achieved
if we consider a Hamiltonian with a SU(2) $\otimes$
SU(2) symmetry rather than the simple SU(2) symmetry of
(\ref{H1}):
\be
H={\bf \Omega}\cdot \bigl[\hat{\bf J}^{(1)}-\hat{\bf J}^{(2)}\bigr].\label{H2}
\ee
Here the angular moments fulfill the commutation relations
\begin{equation}
[{\hat J}^{(n)}_i ,{\hat J}^{(n)}_j]  =  i\varepsilon_{ijk} 
{\hat J}_k^{(n)},\qquad [{\hat J}^{(1)}_i ,{\hat J}^{(2)}_j]  =  0,
\end{equation}
which is equivalent to the Lorentz group.
Staring again with ${\bf \Omega}(0)=(0,0,\Omega)$
and in an eigenstate of (\ref{H2}),  we find that the condition
for a vanishing dynamical phase reads
\be
H|\psi_0\rangle = \hbar\Omega \Bigl[\hat J_z^{(1)}-\hat J_z^{(2)}\Bigr]
\, |\psi_0\rangle =0,
\ee
while the geometric phase is determined by
\be
 {\rm e}^{-i\gamma_{\rm g}} |\psi_0\rangle
=  {\rm e}^{-i\frac{\pi}{2} 
\bigl(\hat J_z^{(1)}+\hat J_z^{(2)}\bigr)}|\psi_0\rangle.
\ee
Since $[\hat J_z^{(1)}+\hat J_z^{(2)},\hat J_z^{(1)}-\hat J_z^{(2)}]=0$
there is the possibility of finding a zero-eigenvalue state of $H$ 
with a geometric phase equal to a multiple of $\pi/2$. 
The existence of
such a state depends on the realization of the Hamiltonian (\ref{H2})
in particular on the total spins $J^{(1)}$ and $J^{(2)}$.

\begin{figure}[ht] 
  \begin{center} 
    \includegraphics[width=6cm]{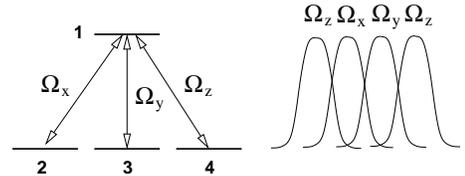} 
    \caption{{\it left:} Tripod coupling scheme representing Hamiltonian
(\ref{H2}) with SU(2) $\otimes$ SU(2) dynamical symmetry.
{\it right:} Pulse sequence to generate geometric phase without dynamical
phase.
}
    \label{Fig2} 
  \end{center} 
\end{figure}

It is easy to see that the simplest possible realization is that of two
half spins, i.e. $J^{(1)}=J^{(2)}=1/2$. If we take into account that
${\bf\Omega}=\{\Omega_x,\Omega_y,\Omega_z\}$ has three independent
components, the $J^{(1)}=J^{(2)}=1/2$ realization of (\ref{H2}) 
corresponds to a four-level scheme coupled by three coherent fields. 
The spin operators can then be expressed e.g. by the following 
$4\times 4$ matrices
\begin{eqnarray}
\hat J_x^{(1)} &=& \frac{1}{2}
\left[\begin{array}{cccc}0 & 1 & 0 & 0\\
              1 & 0 & 0 & 0\\
              0 & 0 & 0 & -i\\
              0 & 0 & i & 0\end{array}\right]\quad
\hat J_x^{(2)} = \frac{1}{2}
\left[\begin{array}{cccc}0 & -1 & 0 & 0\\
              -1 & 0 & 0 & 0\\
              0 & 0 & 0 & -i\\
              0 & 0 & i & 0\end{array}\right]\nonumber\\
\hat J_y^{(1)} &=& \frac{1}{2}
\left[\begin{array}{cccc}0 & 0 & 1 & 0\\
              0 & 0 & 0 & i\\
              1 & 0 & 0 & 0\\
              0 & -i & 0 & 0\end{array}\right]\quad
\hat J_y^{(2)} = \frac{1}{2}
\left[\begin{array}{cccc}0 & 0 & -1 & 0\\
              0 & 0 & 0 & i\\
              -1 & 0 & 0 & 0\\
              0 & -i & 0 & 0\end{array}\right]\nonumber\\
\hat J_z^{(1)} &=& \frac{1}{2}
\left[\begin{array}{cccc}0 & 0 & 0 & 1\\
              0 & 0 & -i & 0\\
              0 & i & 0 & 0\\
              1 & 0 & 0 & 0\end{array}\right]\quad 
\hat J_z^{(2)} = \frac{1}{2}
\left[\begin{array}{cccc}0 & 0 & 0 & -1\\
              0 & 0 & -i & 0\\
              0 & i & 0 & 0\\
              -1 & 0 & 0 & 0\end{array}\right]\nonumber
\end{eqnarray}
With this the Hamiltonian (\ref{H2}) has the matrix form
\be
H=
\left[\begin{array}{cccc}0 & \Omega_x & \Omega_y & \Omega_z\\
              \Omega_x & 0 & 0 & 0\\
              \Omega_y & 0 & 0 & 0\\
              \Omega_z & 0 & 0 & 0\end{array}\right],
\ee
which represents the tripod scheme introduced in \cite{Unanyan-OptComm-1998}
and shown in Fig.\ref{Fig2}, which has been discussed in the context
of non-Abelian geometric phases in 
\cite{Unanyan-PRA-1999,Duan-Science-2001}. The rotation of ${\bf \Omega}$
in parameter space from the $z$-direction to $x$ 
to $y$ and back to $z$ discussed above corresponds
to a sequence of pulses $\Omega_z\to\Omega_x\to\Omega_y\to\Omega_z$ 
also shown
in Fig.{\ref{Fig2}}. The adiabatic rotation 
in parameter space with solid angle $\pi/2$
can be implemented in a very robust way by this pulse sequence.
The only requirements are sufficiently long pulses for adiabaticity 
and overlap of only consecutive pulses to guarantee a solid angle 
of $\pi/2$. Actual shape, precise timing and amplitude of the pulses
are irrelevant. 

%%%%%%%%%%%%%%%%%%%%%%%%%%%%%%%%%%%%%%%%%%%%%%%%%%%%%%%%%%%%%%%%%%%%%%%%%%

%%%%%%%%%%%%%%%%%%%%%%%%%%%%%%%%%%%%%%%%%%%%%%%%%%%%%%%%%%%%%%%%%%%%%%%%%%

The orthogonal zero-eigenvalue states 
of the interaction Hamiltonian at the initial time
$(\hat J_z^{(1)}-\hat J_z^{(2)})|\psi\rangle =0$ are obviously levels
$|2\rangle$ and $|3\rangle$. Their coherent superpositions
%
%
%\be
$\frac{1}{\sqrt{2}}\bigl[|3\rangle \pm i|2\rangle\bigr]$
%\ee
%
%
are also eigenstates of $\hat J_z^{(1)}+\hat J_z^{(2)}$ with eigenvalues
$\pm 1$. After the sequence of four pulses these states acquire
a geometric phase shift of
%
%
%\be
$\gamma_{\rm g}=\mp \frac{\pi}{2}$,
%\ee
%
%
while there is no dynamical phase shift. Applying the pulse
sequence twice leads to a geometric phase of
%
%
%\be
$\gamma_{\rm g}= \pi$.
%\ee
%
%
The latter is also true if the initial state is level $|2\rangle$.
Although $|2\rangle$ is not an eigenstate 
of $U_g$ it is an eigenstates
of $U_g^2$ with eigenvalue $-1$. The last case can also be understood in 
a much simpler way. If initially state $|2\rangle$ is populated, the
double pulse sequence $\Omega_z \to \Omega_x \to \Omega_y \to \Omega_z \to
\Omega_x \to \Omega_y \to \Omega_z$ corresponds to 3 successive 
population transfers via Stimulated Raman adiabatic passage (STIRAP)
\cite{Vitanov-AdvAMO-2001}. In this process
the state vector is rotated according to $|2\rangle\to -|4\rangle\to
+|3\rangle\to -|2\rangle$ thus acquiring a robust phase shift
of $\pi$. 

\

\begin{figure}[ht] 
  \begin{center} 
    \includegraphics[width=7.0cm]{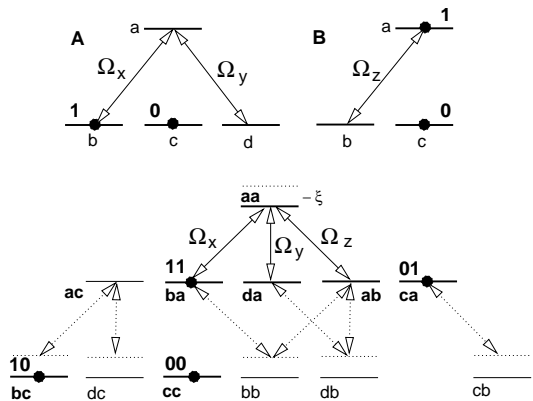} 
    \caption{System of two interacting spin systems.
{\it top:} Spectrum without dipole-dipole interaction. 
$\Omega_x$, $\Omega_y$ and $\Omega_z$ denote coherent couplings. 
{\it bottom:} 
Spectrum of two-particle states with dipole-dipole interaction of strength 
$\xi$. Frequencies of coherent fields are chosen such that the states 
$|{\bf ba}\rangle$,  $|{\bf da}\rangle$,  $|{\bf ab}\rangle$, and  
$|{\bf aa}\rangle$ form a closed tripod system. Logical states 00, 01, 10, and 11
are indicated.
}
    \label{dipole} 
  \end{center} 
\end{figure}

The use of geometric phases in 
the tripod scheme of Fig.~\ref{Fig2} to generate robust single qubit
phase shifts was first suggested by Duan, Cirac and Zoller
in \cite{Duan-Science-2001}. We now show that a similar scheme can be
used to create a robust phase gate between two qubits. 
Here it is important that all states that will be temporarily occupied
are energetically degenerate to avoid
uncontrollable phase shifts if the interaction parameters are
not precisely known. To this end we consider two 
systems with a level structure as shown in Fig.\ref{dipole}
with an interaction 
of the form
\begin{eqnarray}
H &=&H_{0}+H_{dd}+H_{int},  \label{Hamiltonian} \\
H_{0} &=&\mu \sum_{i=1,2}
|a\rangle_{ii}\langle a|,\\
H_{dd} &=&-\hbar\xi |a\rangle_{AA}\langle a| \otimes 
|a\rangle_{BB}\langle a|
,\label{Hdd}\\
H_{int} &=&\hbar\Omega _{x}\Bigl(|a\rangle_{11}\langle b|+h.c.
%|b\rangle_{11}\langle a|
\Bigr) +
\hbar\Omega _{y}\Bigl(|a\rangle_{11}\langle d|+h.c.
%|c\rangle_{11}\langle b|
\Bigr)
+\nonumber\\
&&+
\hbar\Omega _{z}\Bigl(|a\rangle_{22}\langle b|+h.c.
% |b\rangle_{22}\langle a|
\Bigr).
\label{H3}
\end{eqnarray}
$\xi $ represents the strength of the interaction between the spins
in the internal state $|a\rangle$.
An interaction Hamiltonian of type (\ref{H3}) could e.g. be realized with
a pair of atoms with states $|a\rangle$ being Rydberg levels
with a large permanent dipole moment \cite{Jaksch-PRL-2000}.
Also a realization in NMR systems
\cite{NMR} is feasible.
In each of the two systems a single qubit is encoded as indicated in
Fig.~\ref{dipole}:
\be
\bigl [|0\rangle,|1\rangle\bigr]_A
\equiv \bigl[|c\rangle,|b\rangle\bigr]_A,\quad
\bigl[|0\rangle, |1\rangle\bigr]_B \equiv \bigl[|c\rangle,|a\rangle_B
\bigr].\nn
\ee
In order to implement a robust phase gate the interaction shall generate
a geometric phase shift of $\pi$  of state $|ba\rangle$ without 
dynamical phases at all times.

%%%%%%%%%%%%%%%%%%%%%%%%%%%%%%%%%%%%%%%%%%%%%%%%%%%%%%%%%%%%%%%%%%%%%%%%%%

%%%%%%%%%%%%%%%%%%%%%%%%%%%%%%%%%%%%%%%%%%%%%%%%%%%%%%%%%%%%%%%%%%%%%%%%%%

In the lower part of Fig.~\ref{dipole}
the dressed-state energy diagram of the system corresponding to
the free Hamiltonian $H_0$ and the dipole-dipole interaction $H_{dd}$
are shown. Without coherent coupling, i.e. $\Omega_i\equiv 0$
the qubit states $|ca\rangle$ and $|ba\rangle$ will acquire a phase
e$^{-i\mu t}$, while the other two states 
$|cc\rangle$ and $|bc\rangle$ remain constant. The phase
e$^{-i\mu t}$ is without consequence, since it can easily be compensates
by local operations if the energy splitting
$\mu$ is known very well. No precise knowledge of the interaction parameters
$\Omega_i$ is required for this. It is also sufficient to know
the dipole-dipole shift $\xi$ only approximately in order to 
tune the fields close to resonance with $|aa\rangle$.
If $\xi$ is sufficiently large
the coupling
between the states $|ba\rangle$, $|da\rangle$, and $|ab\rangle$
with the lower lying states $|db\rangle$ and $|bb\rangle$ can be disregarded.
In this case there is a tripod coupling with energetically degenerate
states  $|ba\rangle$, $|da\rangle$, and $|ab\rangle$ and
a sequence of overlapping pulses $\Omega_z \to \Omega_x \to  \Omega_y
\to \Omega_z \to \Omega_x \to \Omega_y \to \Omega_z$ will
lead to a geometric phase shift
%
%
%\be
$|{ba}\rangle \, \to\, -|ab\rangle\,
\to\, |da\rangle\, \to\,
-|ba\rangle.
$ 
%\ee
%
%
Since the states  $|ba\rangle$, $|da\rangle$, and $|ab\rangle$ are degenerate
and the interaction $H_{int}$ has SU(2) $\otimes$ SU(2)
symmetry, there is no dynamical phase at any time of the process.
Therefore temporal fluctuations of the field amplitudes and the dipole-dipole
shift will not affect the phase gate. 
It should be noted that the degeneracy of the lower states in the tripod
scheme is important however, since the exact time which the system
spends in the three states depends on the details of the interaction
and is in general not known precisely.

As can be seen from Fig.~\ref{dipole} the off-resonant coupling
of $|ba\rangle$, $|da\rangle$, and $|ab\rangle$ with the lower states
$|db\rangle$ and $|bb\rangle$ can give rise to real transitions 
and ac-Stark dynamical phases. To avoid real transitions into the lower
manifold of states $|\xi|\gg |\Omega_i^{\rm max}|$ is required. The ac-Stark
induced phase shift is negligible if $|\Omega_i^{\rm max}|^2 T /|\xi| \ll 1$
where $T$ is the characteristic time of the process. 
These conditions combined read
\be
\frac{|\Omega_i^{\rm max}|}{|\xi|}\ll 
\frac{|\Omega_i^{\rm max}|^2}{|\xi|} T \ll 1.\label{conditions}
\ee
It should be noted that the dipole-dipole shift $\xi$ is here an
independent parameter and can in principle be chosen very large
without affecting the resonant couplings. This is in contrast to the
proposal of \cite{Duan-Science-2001} where the ac-Stark shifts
cannot be neglected and need to be compensated.

%%%%%%%%%%%%%%%%%%%%%%%%%%%%%%%%%%%%%%%%%%%%%%%%%%%%%%%%%%%%%%%%%%%%%%%%%%

\begin{figure}[ht]
  \begin{center} 
    \includegraphics[width=7.0cm]{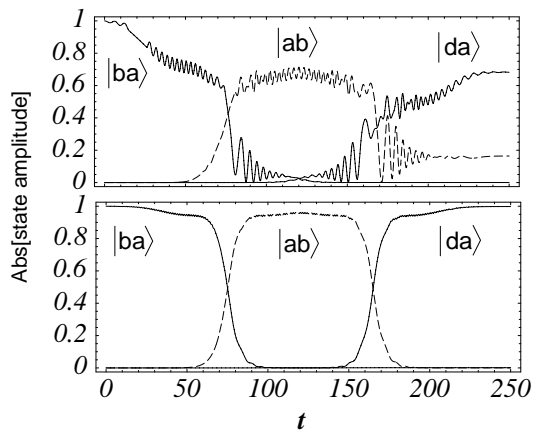} 
    \caption{ Absolute value of state amplitude for 
$|ba\rangle$, $|ab\rangle$, and $|da\rangle$ obtained from
numeric solution of full problem with sequence of 
time-delayed Gaussian pulses $\Omega_z \to \Omega_x\to \Omega_y\to 
\Omega_z$ (half cycle of  phase gate). 
$\Omega_i=\alpha \exp\bigl[-(t-t_i)^2/(2 T^2)\bigr]$ with $\alpha=1$,
$T=20$, $t_x=100$, $t_y=140$, and $t_z=50$ resp. $t_z=190$;
upper curve: $\xi=1$; lower curve: $\xi=4$.
}
    \label{numerics1} 
  \end{center} 
\end{figure} 

%%%%%%%%%%%%%%%%%%%%%%%%%%%%%%%%%%%%%%%%%%%%%%%%%%%%%%%%%%%%%%%%%%%%%%%%%%

\begin{figure}[ht]
  \begin{center} 
    \includegraphics[width=6.5cm]{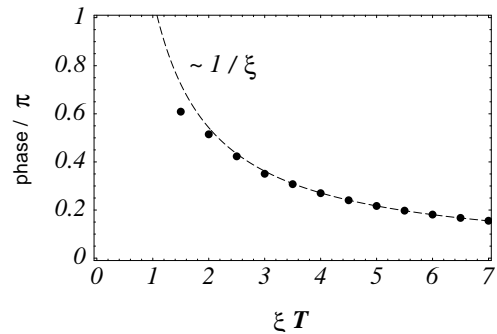} 
    \caption{Final phase of state $|da\rangle$ for pulse sequence 
of Fig.~\ref{numerics1} and
growing values of $\xi$. Dots represent numerical results, dashed curve
is a $1/\xi$ fit.}
    \label{numerics2} 
  \end{center} 
\end{figure}

To illustrate conditions (\ref{conditions}), we 
have numerically calculated the amplitudes of states
$|ba\rangle$, $|da\rangle$, and $|ab\rangle$ as well 
as the phase of the target state $|da\rangle$ for half of the 
phase-gate pulse sequence, i.e. $\Omega_z\to\Omega_x\to\Omega_y\to \Omega_z$. 
The results are shown in Figs.~\ref{numerics1} and \ref{numerics2}. 
Ideally, i.e. if
(\ref{conditions}) is perfectly fulfilled, there is a
complete state transfer from $|ba\rangle$ to $|da\rangle$
with zero phase change. Fig.~\ref{numerics1} 
shows that a dipole-dipole shift slightly larger than the
peak-Rabi frequency is sufficient to suppress real transitions into 
other states. As can be seen from Fig.~\ref{numerics2} the 
ac-Stark induced phase shifts scale 
only as $(\xi T)^{-1}$ and thus larger values of $|\xi|$ are needed
to neglect them. Nevertheless it can be seen that  
it is always possible to choose
sufficiently large values of $\xi$ to obtain a purely geometric phase.

%%%%%%%%%%%%%%%%%%%%%%%%%%%%%%%%%%%%%%%%%%%%%%%%%%%%%%%%%%%%%%%%%%%%%%%%%%

%%%%%%%%%%%%%%%%%%%%%%%%%%%%%%%%%%%%%%%%%%%%%%%%%%%%%%%%%%%%%%%%%%%%%%%%%%

In the present paper we have shown that it is possible to obtain a 
non-vanishing geometric phase of multiples of $\pi/2$ with an exactly
vanishing dynamical phase. For this it is necessary to consider systems
with a SU(2) $\otimes$ SU(2) symmetry rather than just SU(2). The simplest 
nontrivial representation of this symmetry corresponds 
to a four-level system with a tripod coherent coupling. We have shown that
a tripod coupling among two-qubit states can be implemented using 
a pair of coherently driven particles with
dipole-dipole interaction. With this it is possible to design a
geometric phase gate. Due to the absence of dynamical contributions to the 
phase and the
geometric nature of the phase shift, the quantum gate is 
maximally robust against parameter fluctuations. 

\

R.G.U. acknowledges the financial support of the Alexander-von Humboldt
foundation and thanks D. Jaksch for stimulating discussions. This work
is also supported by a grant from the Deutsche Forschungsgemeinschaft
within the Schwerpunktprogramm ``Quanteninformation''.

%%%%%%%%%%%%%%%%%%%%%%%%%%%%%%%%%%%%%%%%%%%%%%%%%%%%%%%%%%%%%%%%%%%

%%%%%%%%%%%%%%%%%%%%%%%%%%%%%%%%%%%%%%%%%%%%%%%%%%%%%%%%%%%%%%%%%%%


\begin{references}


\bibitem{Ekert}  A. Ekert and R. Josza, Rev. Mod. Phys., {\bf 68}, 733
(1996).

\bibitem{error-correction} E. Knill, R. Laflame, and W. H. Zurek, 
Proc. Roy. Soc. London A {\bf 454}, 365 (1998).

\bibitem{Vedral-tutorial} see e.g.: V. Vedral, Int. J. of Quant. Inform. 
{\bf 1}, 1 (2003).

\bibitem{Berry}  M. Berry, Proc. R. Soc. Lond, A {\bf 392}, 45 (1984).

\bibitem{Jones-Nature-2000}  J. A. Jones, V. Vedral, A. Ekert, G. Castagnoli, 
Nature, 
{\bf 403}, 869 (1999).

\bibitem{Duan-Science-2001}  L.-M. Duan, J. I. Cirac, P. Zoller, Science,
{\bf 292}, 1695 (2001).


\bibitem{Garcia-Ripoll-PRL-2003} J.J. Garcia-Ripoll and J.I. Cirac,
Phys. Rev. Lett. {\bf 90}, 127902 (2003).


\bibitem{Unanyan-OptComm-1998} R. Unanyan, M. Fleischhauer, 
B. W. Shore, and K. Bergmann, Opt. Comm. {\bf 155}, 144 (1998).

\bibitem{Unanyan-PRA-1999} R. G. Unanyan, B. W. Shore, and K. Bergmann
      Phys. Rev. A {\bf 59}, 2910-2919 (1999).


\bibitem{Wilczek}  F. Wilczek and A. Zee, Phys. Rev. Lett. {\bf 52}, 2111
(1984).

\bibitem{Aharonov}  Y. Aharonov and J. Anandan, Phys. Rev. Lett. {\bf 58},
1593 (1987).

% \bibitem{Anandan}  J.vAnandan, J. Christian and K. Wanelik, Am. J. Phys. 
% {\bf 65}, 180 (1996).



% \bibitem{Theuer}  H.Theuer, R.G. Unanyan, C. Habscheid, K. Klein and K.
% Bergmann, Optics Express, {\bf 4}, 77 \ (1999).




\bibitem{Robbins}  J.M. Robbins and M. V. Berry, J. Phys. A, {\bf 27}, L435
(1994).



\bibitem{Vitanov-AdvAMO-2001} see e.g.: N.V. Vitanov, M. Fleischhauer, 
B. W. Shore, and K. Bergmann,
Adv. Atom. Mol. Phys.  Vol.{\bf 46}, 55-190 
(ed. by B. Bederson and H. Walther, Academic Press, 2001).

\bibitem{Jaksch-PRL-2000} D. Jaksch, J. I. Cirac, P. Zoller, S. L. Rolston,
R. Cote, and M. D. Lukin, Phys. Rev. Lett. {\bf 85}, 2208 (2000).

\bibitem{NMR} see e.g.: C. P. Slichter, {\it ``Principles of Magnetic Resonance''}
(Springer, Berlin, 1990).



\end{references}
\end{document}